\documentclass[12pt]{article}

\newcommand{\bbr}{I\!\! R}

\newcommand{\x}{arXiv:}
\newcommand{\m}{\mathrm}

\usepackage{graphicx}

\begin{document}
\thispagestyle{empty}
\begin{center}

\null \vskip-1truecm \vskip2truecm

{\Large{\bf \textsf{Fragile Black Holes}}}

{\large{\bf \textsf{}}}

{\large{\bf \textsf{}}}

\vskip1truecm

{\large \textsf{Brett McInnes
}}

\vskip1truecm

\textsf{\\  National
  University of Singapore}

\textsf{email: matmcinn@nus.edu.sg}\\

\end{center}
\vskip1truecm \centerline{\textsf{ABSTRACT}} \baselineskip=15pt
\medskip

The AdS/CFT correspondence may give a new way of understanding field theories in extreme conditions, as in the quark-gluon plasma phase of quark matter. The correspondence normally involves asymptotically AdS black holes with dual field theories which are defined on locally \emph{flat} boundary spacetimes; the implicit assumption is that the distortions of spacetime which occur under extreme conditions do not affect the field theory in any unexpected way. However, AdS black holes are [to varying degrees] \emph{fragile}, in the sense that they become unstable to stringy effects when their event horizons are sufficiently distorted. This implies that field theories on curved backgrounds may likewise be unstable in a suitable sense. We investigate this phenomenon, focussing on the ``fragility" of AdS$_5$ black holes with flat event horizons. We find that, when they are distorted, these black holes are always unstable in string theory. This may have consequences for the detailed structure of the quark matter phase diagram at extreme values of the spacetime curvature.

\newpage

\addtocounter{section}{1}
\section* {\large{\textsf{1. Strongly Coupled Field Theories on Non-Trivial Backgrounds}}}
Strongly coupled field theories are not well understood at very high temperatures or at large values of the field-theoretic chemical potential \cite{kn:gubserreview}. The AdS/CFT correspondence, because it is a strong/weak coupling duality, offers hope of approaching such systems through a dual formulation in terms of semi-classical gravitational configurations in asymptotically AdS spacetimes.

In particular, one wishes to use these methods to understand the \emph{quark matter phase diagram}. The precise structure of this diagram is not known at present; Figure 1 gives one widely accepted picture.
\begin{figure}[!h]
\centering
\includegraphics[width=1.1\textwidth]{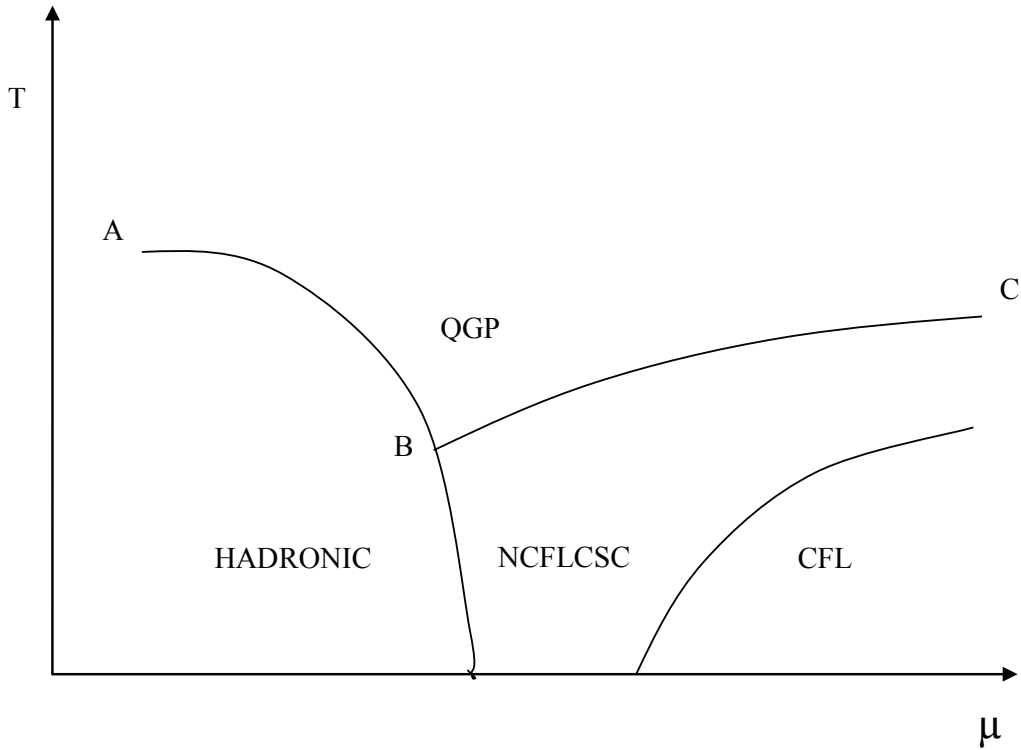}
\caption{Schematic Quark Matter Phase Diagram [after Alford et al. \cite{kn:alford}]; CFL = Colour-Flavour-Locked Phase; NCFLCSC = Non-CFL Colour SuperConductor.}
\end{figure}
The best-understood part of this diagram from a holographic point of view is the high-temperature region, corresponding to the \emph{quark-gluon plasma} or QGP. This is represented by a supersymmetric field theory which is dual to an AdS black hole \cite{kn:chamblin}.

Initially one is concerned with thermal field theories defined on \emph{flat} four-dimensional spacetimes, so the dual system is usually taken to be an asymptotically AdS$_5$ black hole with an exactly flat three-dimensional event horizon. In applications involving the most extreme temperatures and baryon densities, however, the field theory excitations propagate on a background \emph{which is by no means exactly flat}. Consider for example the study of matter in neutron stars \cite{kn:alford2}, where it is possible that the chemical potential of the field theory used to represent QCD may be so high that deconfinement may occur, in a way that one might hope to understand by studying phase transitions as the temperature drops during the formation of the star. In this case, the spacetime curvature cannot be neglected: for example, one must certainly use the full relativistic TOV equation to study the pressure profile of a realistic neutron star. In fact, the curvature of the spacetime in this case is more than ten orders of magnitude larger than in our solar system \cite{kn:psaltis}. Similarly, the most extreme temperatures occur in the earliest history of the Universe, and at sufficiently early times quark matter must have been in the QGP state; again, however, the spacetime during that era is not well approximated by Minkowski spacetime. Thus we see that the regions of Figure 1 that are either very high on the T axis, or just above the line BC, are frequently associated with non-trivial spacetime curvature.

In short, it is exactly in those situations where the AdS/CFT approach is most needed that the flat-space background approximation for the thermal field theory is least satisfactory. It therefore becomes of interest to try to understand the nature of strongly coupled field theories on backgrounds which are not exactly flat, though they may not be far from flatness. The question then becomes: does the introduction of spacetime curvature affect the relevant field theories in any novel or substantial way?

This is a question which can be addressed directly, but one hopes to obtain additional insight by searching for an AdS/CFT dual description. That it is possible to allow non-trivial spacetime geometry at infinity in the AdS/CFT correspondence was discussed in detail by Compere and Marolf \cite{kn:compere}, and this approach has even been successfully extended to dynamical [including FRW] boundary spacetimes \cite{kn:tetrad}\cite{kn:siop}\footnote{Note however that FRW spacetimes are locally \emph{conformally} flat \cite{kn:iihoshi}, and this fact does play a role in the arguments of \cite{kn:tetrad}. Our objective here is to understand the non-conformally-flat case.}. It has also been used to gain potentially crucial new insights into black hole evaporation in the case of strong coupling \cite{kn:hub}; in this case one actually puts a black hole on the boundary.

Here we shall investigate non-trivial geometry at infinity in a simpler way. We shall not allow the spacetime at infinity to be dynamical, nor shall we deform it to the extent considered in \cite{kn:hub}: instead we regard it as a fixed background for the field theory, continuously deformed away from [conformal] flatness. A simple way to proceed in that direction is to consider AdS black holes with event horizons which are not exactly flat. This will correspond to deforming the spatial sections of the boundary spacetime. Our question is then: what happens to an asymptotically AdS black hole, with a flat event horizon, when that event horizon is distorted non-conformally? Ultimately, for realistic distortions, the answers to such questions will require highly sophisticated numerical techniques, as in \cite{kn:cardoso1}\cite{kn:cardoso2}; but we shall see that much can be learned even from toy models based on very simple deformations of the event horizon\footnote{All black holes in this work will be vacuum AdS holes. For studies of AdS black holes with vector ``hair", which also have distorted event horizons, see \cite{kn:pallab}\cite{kn:erd}\cite{kn:gub}. Our results do not extend to this case in a simple way.}.

In the case of AdS black holes with initially \emph{spherical} event horizons, simple distortions of the event horizon have been discussed in the very remarkable work of Murata et al. \cite{kn:murata}; see also \cite{kn:kunz}. It was found that if the sphere which the field theory inhabits is ``squashed" [in a particular way] beyond a certain point, then the theory ceases to be stable. Presumably this means that the corresponding black hole likewise ceases to be stable if its event horizon is similarly squashed\footnote{This instability is in no way related to the questions about stability arising in the study of ``squashed Kaluza-Klein black holes". For the latter, see \cite{kn:harvey}.}. The nature of this instability was not identified in \cite{kn:murata}, but we shall explain it here: it is due to a strictly stringy pair-production phenomenon first discussed by Seiberg and Witten \cite{kn:seiberg} [see also \cite{kn:wittenyau}]. We can say that spherical AdS black holes are somewhat ``\emph{fragile}" in this sense: \emph{if they are deformed excessively then they cease to be stable string-theoretic objects}. These potentially very important results are discussed in Section 2.

Turning to the case of principal interest, black holes with \emph{flat} event horizons, one finds that the results from the spherical case do not carry over in a straightforward way: the Einstein equations with a negative cosmological constant have no solutions corresponding to a black hole with a distorted torus as event horizon. Instead, one would need to consider higher-order corrections to the Einstein equation such as arise in string theory ---$\,$ see for example \cite{kn:odno}\cite{kn:myers}\cite{kn:ohta}. [In this sense, ``squashed flat black holes" are \emph{intrinsically stringy objects.}] Finding a solution of these equations is a formidable exercise indeed, even using numerical methods. Fortunately, deep mathematical results on the space of metrics on tori allow us to make progress here. They imply that the flat case is indeed fundamentally different from the case with spherical event horizons: \emph{all} non-trivial deformations, of whatever magnitude, tend to render the field theory, and the dual black hole, unstable. That is, black holes with flat event horizons are \emph{extremely} ``fragile". These results are the subject of Section 3.

These conclusions are apparently rather alarming, suggesting as they do that field theories defined on flat spacetimes immediately become unstable when the background is distorted non-conformally. However, instability need not be a threat if it evolves sufficiently slowly \cite{kn:albion}\cite{kn:silverymoon}; it can sometimes be interpreted rather as metastability \cite{kn:triple}\cite{kn:barbon}. On the other hand, we do \emph{not} want the black hole description of quark matter to be valid at low temperatures, since it is clear from Figure 1 that the QGP cannot exist under those conditions. The consequences of black hole fragility in this context are discussed in Section 4 and in the Conclusion. The upshot is that fragility causes the precise position of the line BC in Figure 1 to depend on the spacetime curvature. In particular, the temperature of the quark matter triple point [point B] may be higher when the curvature is relatively large, as for example in the region in which a neutron star is formed.
\addtocounter{section}{1}
\section* {\large{\textsf{2. Fragility of Spherical AdS$_5$ Black Holes}}}
Murata et al. \cite{kn:murata} consider \emph{N} = 4 super-Yang-Mills theory defined on a ``squashed Einstein static universe", with metric\footnote{Field theories defined on squashed spheres have been studied extensively: see for example \cite{kn:emp}\cite{kn:porrati}\cite{kn:har}. In those works, however, the bulk does not contain a black hole: it is typically a Taub-NUT space. Here, by contrast, we are concerned exclusively with thermal field theories on the boundary, dual to asymptotically AdS black holes in equilibrium with their own Hawking radiation.}
\begin{equation}\label{A}
\m{g(SqESU) = -\,dt^2\;+\;{L^2\over 4}\,\Big[\,(\sigma^1)^2\;+\;(\sigma^2)^2\;+\;s^2\,(\sigma^3)^2 \Big]}.
\end{equation}
Here $\sigma^{1,2,3}$ are the usual orthogonal basis one-forms on the unit three-sphere S$^3$, L is a constant with units of length, and s is a dimensionless constant non-zero ``squashing" parameter whose deviation from unity measures the extent of the deformation of the spatial sections away from the canonical ``round" shape. Of course, this is not an interesting background for applications to realistic field theories, but we shall see that it [and even more so its initially spatially flat counterparts, to be discussed in the next section] does exhibit some crucial behaviour which turns out to be generic.

The scalar curvature of this metric is non-zero only because of the curvature of the spatial sections ---$\,$ in fact this will be true of all of the boundary metrics we shall consider here ---$\,$ and is given by
\begin{equation}\label{B}
\m{R[g(SqESU)]\;=\;{2\over L^2}\,\Big[\,4\;-\;s^2 \Big].}
\end{equation}
We see that this particular kind of deformation increases the scalar curvature if s $<$ 1 [the oblate spheroids], but decreases it if s $>$ 1 [prolate spheroids]. Crucially, we see that for sufficiently prolate spheroids, that is, for
sufficiently large s, the scalar curvature of a topological three-sphere can be \emph{constant and negative}; this, as is clear from the Gauss-Bonnet theorem, would not be possible for a two-sphere. Notice that intuitions shaped by the two-dimensional case are likely to be misleading here; henceforth, all event horizons under discussion have dimension at least three.

Murata et al. find that the scalar sector of the theory becomes tachyonic when s $>$ 2, that is, precisely when the squashing becomes so extreme that the scalar curvature is negative. Naively, this is to be expected, because the usual conformal coupling term acts like a negative squared mass term in this case. Thus the theory is ill-defined if the boundary is deformed to this extent\footnote{Murata et al. also find that the spinor sector of the theory becomes tachyonic, but only when s $>$ 4.}. [The actual situation is considerably more subtle, but we defer a consideration of this issue until Section 4, below.] Note that not all deformations of the three-sphere tend to render the field theory tachyonic: oblate deformations are perfectly safe in that sense [though certain other kinds of misbehaviour, which we shall not discuss here, arise in the extremely oblate case \cite{kn:murata}.] The generality of this result, and of those to be discussed below, strongly suggests that similar conclusions hold in more realistic thermal field theories, and as usual we shall assume henceforth that this is the case.

Murata et al. also find the bulk geometry which has an Einstein metric [of negative cosmological constant], which contains a black hole, and which has a boundary geometry of the type given by equation (\ref{A}). A ``round" Schwarzschild black hole in AdS$_5$ has a metric of the form
\begin{eqnarray}\label{CC}
\m{g(\Lambda < 0;\;Sch; S^3)} &=& \m{- \Bigg[{r^2\over L^2}\;+\;1\;-\;{8 M\over
3 \pi r^2}\Bigg]dt^2\;+\;{dr^2\over {r^2\over L^2}\;+\;1\;-\;{8
M\over 3 \pi r^2}}} \\
&&\;+\; \m{{r^2\over 4}\,\Bigg[\,(\sigma^1)^2\;+\;(\sigma^2)^2\;+\;(\sigma^3)^2 \Bigg]}. \nonumber
\end{eqnarray}
Here L is the radius of curvature of the asymptotic AdS$_5$, M is the ADM mass, the notation S$^3$ reminds us that the event horizon has the topology of a sphere, and $\sigma^{1,2,3}$ are as above. In order to find the metric of the AdS$_5$ ``squashed Schwarzschild" black hole, we need to consider Einstein metrics [with $\Lambda < 0$] which are
asymptotically AdS$_5$: these take the form
\begin{eqnarray}\label{C}
\m{g(\Lambda < 0;\;SqSch; S^3)} \;=\;\m{-\,F(r)e^{-2\delta (r)}\,dt^2 \;+\;{dr^2\over F(r)}}
\;+\;\m{{r^2\over 4}\,\Bigg[\,(\sigma^1)^2\;+\;(\sigma^2)^2\;+\;s(r)^2\,(\sigma^3)^2 \Bigg].}
\end{eqnarray}
Here F(r) is a non-negative function, asymptotically of order r$^2$, the vanishing of which defines the location of the event horizon; $\delta$(r) is a bounded function, vanishing at infinity, which is needed \cite{kn:jacobson} if one is to find a solution in the case where the event horizon is not a space of constant curvature; and s(r) is a function which can have an asymptotic positive value s$_{\infty}$ not necessarily equal to unity, as in equation (\ref{A}). We have here a two-parameter family of metrics; one can take these parameters to be the values of r and of s(r) at the horizon. It is the fact that the field equations and regularity conditions \emph{do not} fix this latter parameter that makes it possible to find a black hole dual to a field theory on any squashed three-sphere.

The precise forms of F(r), $\delta$(r), and s(r) must of course be found numerically: see \cite{kn:murata} for the details. In particular, one finds that the transverse sections r = constant are spacetimes with spatial sections having the same character as the spatial sections at infinity: if the latter are for example prolate, then all of the r = constant, t = constant sections are prolate squashed spheres. In every case except that of black holes with perfectly spherical spatial sections at infinity [s$_{\infty}$ = 1], the degree of squashing \emph{decreases} as one moves from infinity towards the event horizon; that is, s(r) is monotonically decreasing in the oblate case, but monotonically increasing in the prolate case. However, the value of s(r) at the event horizon, s$_{\m{H}}$, is not equal to unity unless s$_{\infty}$ = 1. The physical intuition here is that the distorted shape of the event horizon is sustained by still stronger deformations imposed at the boundary. This is how one understands the fact that s(r) cannot be a constant except in the trivial case where there is no deformation either at the event horizon or at infinity.

In the critical case where the field theory at infinity is on the brink of becoming tachyonic [that is, s$_{\infty}$ = 2], one finds, if the value of r at the event horizon, r$_{\m{H}}$, is equal to unity, that
\begin{equation}\label{D}
\m{s_{H}\;\approx \; 1.19 \;\;\;\;\;\;[\,s_{\infty}\;=\;2,\; r_H = 1.]}
\end{equation}
This represents a rather mild degree of squashing at the event horizon. Furthermore, it follows from Figure 2 of \cite{kn:murata} that s$_{\m{H}}$ becomes still smaller [for s$_{\infty}$ fixed at 2] as r$_{\m{H}}$ is reduced; it approaches unity. As the area of the event horizon is 2$\pi^2$r$_{\m{H}}^3$s$_{\m{H}}$, this means that, for black holes with small masses and areas, \emph{the amount of distortion of the event horizon required to induce instability in the boundary theory can be very small}.

Murata et al. stress that all this leads directly to an apparent paradox: the field theory evidently does not exist as a unitary theory when s$_{\infty}$ $>$ 2, yet even in that case there appears to be nothing to prevent the existence of a dual black hole in the bulk with a [not very] prolate event horizon. If such black holes could be shown to exist as stable objects in the full string theory in the bulk, this would constitute a significant failure of AdS/CFT, precisely in a case of potentially great interest in applications. [The situation in the non-thermal case \cite{kn:emp}\cite{kn:porrati}\cite{kn:har} is rather complex, but is thought to be fully compatible with AdS/CFT.]

Happily, while arbitrarily prolate black holes are perfectly acceptable classically, excessively deformed event horizons are a signal of instability in the full theory. For, as we have seen, the spatial sections at infinity can be squashed only if the transverse spheres in the bulk are likewise squashed. This has a subtle effect on the areas and volumes of branes in the bulk, altering the value of the action for BPS branes, particularly at large distances from the black hole. The result can be an infinite reservoir of negative free energy, leading to the form of instability discovered by Seiberg and Witten \cite{kn:seiberg}. As Maldacena and Maoz put it in another context \cite{kn:maoz}, this is a kind of pair-production instability for certain branes.

We are interested in the action of a BPS brane wrapping an r = constant section of the [Euclidean version of] the space with metric given by equation (\ref{C}). This can be computed by means of a straightforward adaptation of the methods given in, for example, \cite{kn:seiberg}\cite{kn:wittenyau}\cite{kn:maoz}\cite{kn:porrati}: it takes the form
\begin{equation}\label{E}
\m{\$(\Lambda < 0;\;SqSch; S^3; s_H, r_H, r) \;=\;4 \pi^3 \Theta P \Bigg[
r^3F^{1/2}s(r)e^{-\,\delta(r)}\;-\;{4\over L}\int_{r_H}^r \rho^3s(\rho)e^{-\,\delta(\rho)}d\rho \Bigg].}
\end{equation}
Here $\Theta$ is the brane tension, r$_{\m{H}}$ and s$_{\m{H}}$ are the parameters of the black hole, and 2$\pi$P is the period of the Euclidean ``time" coordinate [chosen, as always, so that the Euclidean version of the metric is non-singular at r = r$_{\m{H}}$; P can be expressed in terms of r$_{\m{H}}$ and s$_{\m{H}}$]. Thus if r$_{\m{H}}$ and s$_{\m{H}}$ are fixed, we can think of the action as a function of r only.

It is possible to show that the slope of this function is positive near to r = r$_{\m{H}}$, and clearly the action vanishes there; hence the action is certainly non-negative near to the event horizon. However, since $\delta(\m{r})$ vanishes towards infinity and s(r) tends to a constant, we see that both terms inside the brackets in (\ref{E}) diverge quartically as r tends to infinity. It is therefore far from clear that the action remains non-negative for large values of r. One should expect that it will do so for some pairs (r$_{\m{H}}$, s$_{\m{H}}$), but not for others. In the latter case, one can expect uncontrolled pair-production of branes to occur in the corresponding region, and the system will be unstable.

One could investigate this question numerically, seeing how the asymptotic value of the action changes as the pair (r$_{\m{H}}$, s$_{\m{H}}$) varies, but the work described in \cite{kn:seiberg} allows us to side-step that computation. For Seiberg and Witten were able to show very generally that the BPS brane action \emph{must} become negative at large distances ---$\,$ in fact, it is unbounded below ---$\,$ \emph{whenever the scalar curvature at infinity [in the Euclidean version of the theory] is negative}. [Conversely, the action at large distances is positive if the scalar curvature at infinity is positive.] Thus we see that black holes of the form (\ref{C}) do not exist as static objects in string theory when the asymptotic value of s(r) exceeds 2. In short, spherical AdS black holes do indeed become unstable when their event horizons are distorted to a rather small degree. This is the surprising phenomenon of black hole ``\emph{fragility}". It is this non-perturbative effect that saves the AdS/CFT correspondence in this case.

We can summarize the situation as follows. Consider a thermal field theory defined on the Einstein static universe [that is, $\bbr \,\times \,$S$^3$ with the metric (\ref{A})]. Then

\bigskip

$\bullet$ Some deformations of S$^3$, but not all, will ultimately cause the theory to become unphysical.

\bigskip

$\bullet$ If one regards this field theory as being defined on the conformal boundary of a dual AdS black hole spacetime, then this black hole is ``metastable", in the sense that no infinitesimal distortion of the event horizon will induce instability; however, the amount of [prolate] squashing of the event horizon required to do so is quite small, particularly for black holes with a small horizon area. Thus spherical AdS$_5$ black holes become increasingly fragile as they become smaller.

\bigskip

$\bullet$ The above two points are of intrinsic interest, but, additionally, in combination they confirm the validity of the AdS/CFT correspondence. The essential physical phenomenon here is Seiberg-Witten instability \cite{kn:seiberg}. This effect is evidently the key to understanding the behaviour of \emph{any} AdS black hole under distortion.

\bigskip

Only spherical black holes were considered in \cite{kn:murata}, but one expects qualitatively similar conclusions to hold in the physically more interesting case in which the field theory is initially defined on a \emph{flat} spacetime, as opposed to the Einstein static universe we have considered thus far. The actual results, to which we now turn, therefore come as a surprise.

\addtocounter{section}{1}
\section* {\large{\textsf{3. Fragility of Flat AdS$_5$ Black Holes}}}
A black hole is a localized object with an interior fenced off by its event horizon. Any compact connected space which can be the boundary of a manifold-with-boundary is therefore a candidate for such a surface, but in practice the \emph{dominant energy condition} usually rules out most possibilities \cite{kn:greg1}\cite{kn:greg2}. As asymptotically AdS spacetimes do not satisfy this condition, however, it becomes possible to consider black holes with event horizons having non-spherical topologies. In particular, one can consider $\Lambda < 0$ black holes with compact \emph{flat} event horizons \cite{kn:lemmo}. The conformal boundary can then be regarded as an exactly locally flat manifold.

In more detail, the $\Lambda < 0$ black hole with event horizon having the exact geometry of the flat torus, T$^3$, has the metric
\begin{equation}\label{F}
\m{g(\Lambda < 0;\;Sch; T^3) = - \Bigg[{r^2\over L^2}\;-\;{2M\over
3\pi^2K^3r^2}\Bigg]dt^2\;+\;{dr^2\over {r^2\over L^2}\;-\;{2M\over
3\pi^2K^3r^2}} \;+\; r^2\Big[dx^2\;+\;dy^2\;+\;dz^2 \Big]}.
\end{equation}
Here $-1$/L$^2$ is the asymptotic curvature, M is the ADM mass, and x,y,z are dimensionless Cartesian coordinates on a torus, which we take to be cubic; that is, all three coordinates are periodically identified\footnote{For the need to include the factors involving K in equation (\ref{F}), see \cite{kn:peldan}. One obtains a ``planar" black hole in the limit K $\rightarrow$ $\infty$, but this means that we would have to consider only infinitely massive black holes. We prefer, following for example \cite{kn:kirit}, to retain the flexibility to have a black hole of any mass. If necessary we can take K to infinity in our final results, using the action per unit area instead of the total action when computing brane actions. Spatial compactification is in fact a promising technique for understanding QCD at non-zero chemical potential: see the recent remarkable work of Hands et al. \cite{kn:hands}.} with the same period, 2$\pi$K. Note that the Euclidean version of this metric has, as usual, a periodically identified ``time" coordinate; therefore, the conformal boundary in the Euclidean case has the structure of a four-dimensional torus, which is flat [but not cubic in general]. The topology will of course remain unaffected even if the geometry of the black hole is deformed.

The Seiberg-Witten criterion for stability against brane nucleation, as discussed earlier, has nothing to say here, since the scalar curvature at infinity is evidently zero in this case. However, the action for branes in the Euclidean version of the space with metric given in (\ref{F}) can be computed explicitly, and its graph [for typical parameter values] is shown in Figure 2 [see \cite{kn:conspiracy}].
\begin{figure}[!h] \centering
\includegraphics[width=0.9\textwidth]{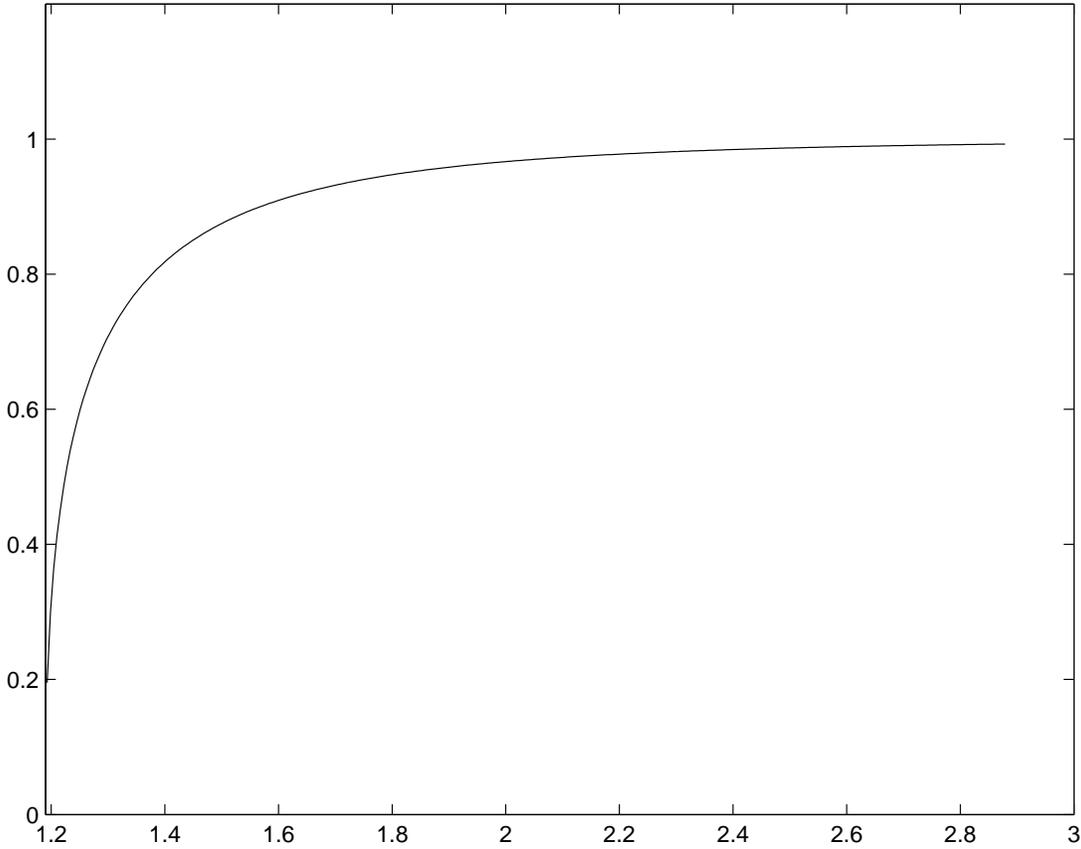}
\caption{Brane Action, Exactly Flat Event Horizon.}
\end{figure}
Clearly the action is positive everywhere; it is asymptotic to a positive constant, which is in fact determined by the entropy of the black hole. Thus the unperturbed flat black hole is certainly stable.

We are interested in determining whether this statement still holds under non-conformal deformations of the event horizon. Of course, there are many possible deformations of this kind: one might begin by considering metrics of the form
\begin{eqnarray}\label{G}
\m{g(\Lambda < 0;\;SqSch; T^3)}\;  &=&  \; -\,\m{F(r)e^{-2\delta (r)}\,dt^2 \;+\;{dr^2\over F(r)}} \\
& & +\;\m{r^2\,\Big[\,A(x,y,z,r)dx^2\;+\;B(x,y,z,r)dy^2\;+\;C(x,y,z,r)dz^2\Big].} \nonumber
\end{eqnarray}
Here A(x,y,z,r), B(x,y,z,r), and C(x,y,z,r) are strictly positive functions which are periodic in x, y, z, with period 2$\pi$K. This form is still too general to be tractable: as in \cite{kn:murata}, we shall consider the simplest possible squashings; so we set A(x,y,z,r) = 1, and take only the lowest Fourier modes of B(x,y,z,r) and C(x,y,z,r). Now if the event horizon is deformed but remains locally conformally flat ---$\,$ the event horizon of a five-dimensional black hole being three-dimensional, this can be detected by means of the so-called Cotton tensor \cite{kn:cotton} ---$\,$ then the hole's four-torus at infinity likewise remains locally conformally flat. The conformally flat case has been considered in \cite{kn:tetrad}, so we are interested in choosing our metric ansatz so as to avoid flat or conformally flat deformations of the event horizon. Essentially this just means that the components of the metric in a given direction on the torus must depend on the coordinates describing the \emph{other} directions. This leads us to a metric of the form
\begin{eqnarray} \label{H}
 \m{g(\Lambda < 0;\;SqSch; T^3; \alpha(r), \beta(r))}& = &\m{-F(r)\,e^{-2\delta (r)}\,dt^2\;+\;{dr^2 \over F(r)}} \; +\;\m{r^2\,\Bigg[\,dx^2\,+}
 \\
&& \m{\Bigg(1 + \alpha(r)cos\Big({x\over K}\Big)\Bigg)^2 dy^2 + \Bigg(1 + \beta(r)cos\Big({y\over K}\Big)\Bigg)^2 dz^2 \Bigg].} \nonumber
\end{eqnarray}
Here $\alpha$(r) and $\beta$(r) are functions which must be constrained to satisfy $|\,\alpha$(r)$| < 1$, $|\,\beta$(r)$| < 1$. [More precisely, they must be bounded away from unity as r tends to infinity.] The objective now is to find an Einstein metric of this form with $\Lambda < 0$. Our experience with the spherical case, and also physical intuition, lead us to expect that the boundary geometry will be ``more deformed" than that of the event horizon; so, apart from cases in which $\alpha$(r) or $\beta$(r) vanish identically, there will be no solutions with constant $\alpha$(r) or constant $\beta$(r). These functions will be determined by  $\alpha_{\m{H}}$ and $\beta_{\m{H}}$, their values at the horizon, which is located at r = r$_{\m{H}}$. The three numbers $\alpha_{\m{H}}$, $\beta_{\m{H}}$, and r$_{\m{H}}$ are the parameters which specify the black hole.

The Euclidean version of the boundary metric will take the form
\begin{equation} \label{HORRIBLE}
 \m{g(Sq T^4; \alpha_{\infty}, \beta_{\infty})\; =\; dt^2\;+\;L^2\,\Bigg[dx^2\,+
 \Bigg(1 + \alpha_{\infty}cos\Big({x\over K}\Big)\Bigg)^2 dy^2 + \Bigg(1 + \beta_{\infty}cos\Big({y\over K}\Big)\Bigg)^2 dz^2 \Bigg],}
\end{equation}
where $\alpha_{\infty}$ and $\beta_{\infty}$ are the asymptotic values of $\alpha$(r) and $\beta$(r); they are fixed by $\alpha_{\m{H}}$ and $\beta_{\m{H}}$.
This is the metric of a squashed four-torus [with only two directions actually squashed in this case]. Of course, such a metric cannot be regarded as the Euclidean version of an approximation to the metric inside, for example, a neutron star; that would be much more complicated, and in particular the coefficient of the dt$^2$ term would be far from trivial. We shall see, however, that even this very simple geometry shares a crucial property with more realistic background spacetimes for real field theories.

Having found a solution of the form given in equation (\ref{H}), we would then wish to analyse the brane action, as in the previous section: here it takes the form\footnote{Readers who prefer to think in terms of ``planar" black holes [that is, holes with actually infinite event horizons] can compute an action per unit area by dividing this expression by the area of a reference brane located at some fixed value of r. This will remain finite as K is taken to infinity.}
\begin{equation}\label{HH}
\m{\$(\Lambda < 0;\;SqSch; T^3; \alpha_H, \beta_H, r) \;=\;16 \pi^4 \Theta P K^3 \Bigg[
r^3F^{1/2}e^{-\,\delta(r)}\;-\;{4\over L}\int_{r_H}^r \rho^3e^{-\,\delta(\rho)}d\rho \Bigg],}
\end{equation}
where the notation is as in equation (\ref{E}). Notice that, as indicated, this does indeed depend on the black hole parameters $\alpha_{\m{H}}$ and $\beta_{\m{H}}$, since these affect the metric component functions F(r) and $\delta$(r). As before, we are particularly interested in the asymptotic behaviour of this function, and, again, its sign at infinity can be investigated by examining the scalar curvature of the metric in (\ref{HORRIBLE}).

This scalar curvature is given by
\begin{equation}\label{HHH}
\m{{2 \Bigg(\beta_{\infty} cos\Big({y\over K}\Big)+\alpha_{\infty}cos\Big({x\over K}\Big)+\alpha_{\infty}^2cos^2\Big({x\over K}\Big)+\beta_{\infty}cos\Big({y\over K}\Big)\alpha_{\infty}cos\Big({x\over K}\Big)+
\alpha_{\infty}^2cos^2\Big({x\over K}\Big)\beta_{\infty}cos\Big({y\over K}\Big)\Bigg)/L^2}\over K^2\Bigg({1+\beta_{\infty}cos\Big({y\over K}\Big)+\alpha_{\infty}^2cos^2\Big({x\over K}\Big)+\alpha_{\infty}^2cos^2\Big({x\over K}\Big)\beta_{\infty}cos\Big({y\over K}\Big)+
2\alpha_{\infty}cos\Big({x\over K}\Big)+2\beta_{\infty}cos\Big({y\over K}\Big)\alpha_{\infty}cos\Big({x\over K}\Big)\Bigg)}}.
\end{equation}
This function is far from being positive everywhere, as may be seen [for typical values of the parameters] in Figure 3.
\begin{figure}[!h]
\centering
\includegraphics[width=0.8\textwidth]{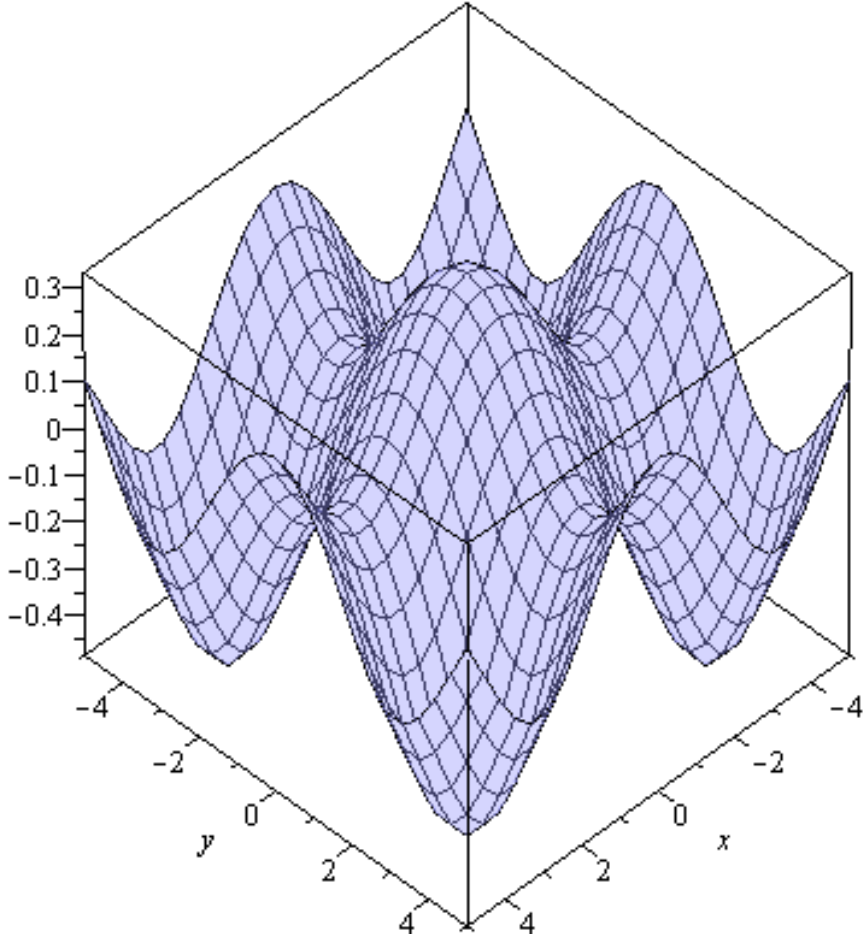}
\caption{Scalar Curvature of $\m{g(Sq T^4; \alpha_{\infty}, \beta_{\infty})}$, K = 1, $\alpha_{\infty}$=$\beta_{\infty} = 0.1$.}
\end{figure}
However, as we shall discuss below, it can be rendered constant by means of a conformal transformation of the metric; since the boundary metric is given only up to a conformal factor in any case, nothing is lost by so doing. The question is whether the constant obtained when the graph is ``flattened out" is positive or negative [or, indeed, zero]. We shall return to this problem below.

Returning to the metric given in (\ref{H}): even in this simplified form, it generates a Ricci tensor of formidable complexity; it can be computed using a programme such as MAPLE, but, as it would run for several pages, we shall not try to reproduce it here. However, an examination of the output reveals the following fact: \emph{the Ricci tensor is not diagonal in general} in these coordinates. For example, the xy component of the Ricci tensor is
\begin{equation} \label{I}
\m{R_{x y}\;=\; {2 \alpha(r)\beta(r)sin\Big({x\over K}\Big)sin\Big({y\over K}\Big)\over K^2 \Big(1 + \alpha(r)cos\Big({x\over K}\Big) + \beta(r)cos\Big({y\over K}\Big) + \alpha(r)\beta(r)cos\Big({x\over K}\Big)cos\Big({y\over K}\Big)\Big)},}
\end{equation}
while the rx component is
\begin{equation} \label{J}
\m{R_{r x}\;=\; {2 {d\alpha(r) \over dr}sin\Big({x\over K}\Big)\over K \Big(1 + \alpha(r)cos\Big({x\over K}\Big)\Big)},}
\end{equation}
and R$_{\m{ry}}$ involves the derivative of $\beta$(r) in a similar way. Thus to obtain an Einstein metric we must set $\beta$(r) = 0 and $\alpha$(r) = constant [or vice versa, to the same effect]. But we argued earlier that there are no such solutions except when $\alpha$(r), too, vanishes identically. [Even if such a solution did exist, the event horizon would be conformally flat, the case we are trying to avoid.]

Thus we have a curious result: for black holes with flat event horizons, there is no analogue of the procedure followed in \cite{kn:murata}. We cannot ``squash" the black hole and observe the consequences, at least not at the level where the bulk metric can be approximated by a metric like that of AdS$_5$ itself, that is, an Einstein metric of negative scalar curvature. Instead, one would have to consider higher-order corrections to the gravitational action in the bulk, as in the ``Gauss-Bonnet" and related actions discussed, for example, in \cite{kn:odno}\cite{kn:myers}\cite{kn:ohta}. In that case, the Ricci tensor need not be diagonal, and so it may well be possible to find non-trivial metrics of the form given in equation (\ref{H}). In terms of the AdS/CFT correspondence, this means that we are forced to deal with field theories such that the 't Hooft coupling and the number of colours are not arbitrarily large \cite{kn:ge}. One could argue that this is not surprising, since this step will have to be taken in any case as one moves towards more realistic field theories on the boundary; however, it is remarkable that merely deforming the background away from flatness should enforce this procedure.

The next step, then, is to insert a metric of the form given in equation (\ref{H}) into the field equations of a Gauss-Bonnet type of bulk theory, and search for non-nakedly-singular [black hole] solutions; one might try to follow, for example, \cite{kn:brandon}. The results, however, are extremely complicated, even using an algebraic computing programme. We shall argue later that, nevertheless, it might be useful to proceed with such a calculation. Here we shall only sketch the general features of the results; fortunately, this can be done very simply with the aid of certain recent results in pure geometry.

We begin with the observation that, whether the event horizon is distorted or not, the topology of the Euclidean version of the conformal boundary remains that of T$^4$. The scalar curvature of the undistorted metric on T$^4$ is of course zero; we need to understand what happens to it when T$^4$ is distorted. Throughout the following discussion, the reader should bear in mind that, in conformal geometry, the scalar curvature can, without loss of generality, be regarded as a constant \cite{kn:schoenyam}.

The first point to make is that there is a strong asymmetry between the cases of the torus and of the sphere. As we have seen, it is perfectly possible to distort the geometry of a sphere in such a way that the scalar curvature becomes a negative constant. But it is \emph{not} possible to distort T$^4$ so that the scalar curvature becomes positive. The [deep and difficult] proof of this fact is due to Schoen, Yau \cite{kn:schoenyau}, Gromov, and Lawson
[\cite{kn:lawson}, page 306]. This has the following physical interpretation in our case. We saw that spherical AdS$_5$ black holes were ``metastable" in the sense that, while they can be rendered unstable by means of [surprisingly small] distortions of their event horizons, they are not on the brink of becoming unstable: no infinitesimal amount of squashing will have this effect. In the case of flat event horizons, however, the situation is worse, for at least two reasons. First, we are already, even before we deform the event horizon, on the brink of having negative scalar curvature at infinity; and, second, we see now that no deformation of the event horizon can drive the system away from this brink.

However, the actual situation is still worse. We are accustomed to the idea that interesting, non-flat spacetimes with vanishing scalar curvature can be found ---$\,$ the usual asymptotically flat Schwarzschild geometry is an obvious example. The point here is simply that requiring the scalar curvature to vanish involves putting only one condition on the curvature tensor, which of course has many independent components; thus it should be easy to satisfy this condition without forcing the space to be locally flat. It is therefore entirely reasonable to hope that, even though there can be no metrics of the form (\ref{H}) with positive scalar curvature at infinity, there might be \emph{many} such metrics with a non-trivial metric at infinity having exactly vanishing scalar curvature. Remarkably, however, this is not the case. Gromov and
Lawson, generalizing a theorem of Bourguignon, proved the following surprising result
[\cite{kn:lawson}, page 308]:

\bigskip
\noindent \textsf{THEOREM [Bourguignon-Gromov-Lawson]: If a metric
on a compact enlargeable spin manifold has zero scalar curvature,
then that metric must be exactly locally flat, that is, the
curvature tensor must vanish identically everywhere on the
manifold.}
\bigskip

\noindent Here ``enlargeability" is a technical topological condition which is often satisfied if the fundamental group of the manifold is infinite; it is satisfied by a torus of any dimension, as also is the ``spin" condition; see \cite{kn:lawson} for the details.

The Bourguignon-Gromov-Lawson theorem, combined with the Schoen-Yau-Gromov-Lawson result discussed earlier, has the following dramatic consequence: \emph{every} non-conformal deformation of a flat torus will result in a metric which, when conformally transformed so that the scalar curvature is constant, then has \emph{negative} scalar curvature. In pictorial terms, we can flatten out the graph in Figure 3 [which portrays the scalar curvature of the metric on T$^4$ given by equation (\ref{HORRIBLE}), which is not conformally flat in general] by means of a conformal transformation: but, when we do so, the flattened function will be a \emph{negative} constant.

The upshot is that, for non-conformal deformations, the action function in equation (\ref{HH}) will \emph{always} ---$\,$ that is, for all non-zero values of the black hole parameters $\alpha_{\m{H}}$ and $\beta_{\m{H}}$ ---$\,$ become negative at some sufficiently large distance from the black hole. In fact, the conclusion is much stronger than this. For the sake of simplicity [and because it is natural from the black hole point of view] we have so far confined ourselves to deformations of the boundary geometry which correspond to deformations of the three-dimensional geometry of the black hole event horizon. Essentially this means that we have only considered deformations of the spatial geometry at infinity. But the results above mean that \emph{all} non-conformal deformations of the geometry at infinity give rise to a brane pair-production instability. Thus, the conclusions we are reaching in these simple toy models are in fact \emph{generic}.

For example, suppose that one wished to consider a realistic interior neutron star spacetime, which might involve the study of a field theory which deconfines due to the extreme conditions in the core \cite{kn:alford2}, associated with very large values of the baryonic chemical potential. One can hope to work towards a holographic account of this system by beginning with a quark-gluon plasma corresponding to a point in the quark matter phase diagram [Figure 1] which lies at a similar value of the chemical potential but which has a higher temperature. The ``quark fluid" might be studied by gradually cooling this plasma.
The dual object describing the plasma is an electrically charged asymptotically AdS black hole with an event horizon which ceases to be flat when the spacetime curvature associated with the star is taken into account. This system is very much more complicated than the black holes we have considered here, yet the above results allow us to predict that it is unstable in the Seiberg-Witten sense. Similarly, in the still more extreme case of a field theory defined on a black hole spacetime, the dual system, discussed in \cite{kn:hub}, involves [at least for small black holes] a deformed planar black hole in the bulk. This system too must give rise to bulk branes with an action function which becomes negative at some point. The application to the cosmological case \cite{kn:tetrad}\cite{kn:siop} works in a different way, since all exact FRW spacetimes are locally conformally flat \cite{kn:iihoshi}. Thus if one focuses on FRW cosmologies with flat compact spatial sections, then the Euclidean version is in fact conformal to an exactly flat four-torus and so there is no instability in this case\footnote{Our Universe appears in fact to have been locally conformally flat, to an extreme degree of accuracy, in its earliest infancy ---$\,$ this observation underlies Penrose's ``Weyl Curvature Hypothesis" [see for example \cite{kn:penrose}\cite{kn:lebowitz}\cite{kn:zeh}]. In the context of toral cosmologies, one might speculate that this very peculiar state of affairs is related to the fact that a non-vanishing Weyl tensor would, according to our discussion, give rise to an instability. See, however, the discussion below regarding thermal effects on the rate at which instability would develop.} However, one would of course still have to take our results into account when dealing with the small perturbations away from exact conformal flatness which have occurred throughout cosmic history.

In summary, then, we have two findings. The first is that ``squashed" flat AdS black holes can only be understood in the context of higher-order corrections to the Einstein equations; they are intrinsically \emph{stringy} objects. The second is that such black holes are always unstable to a particular stringy effect. In short, if we say that spherical AdS black holes are ``fragile" in the sense discussed in the preceding section, then ``flat" AdS black holes can only be described as \emph{extremely} fragile.

\addtocounter{section}{1}
\section* {\large{\textsf{4. Instability vs. Metastability}}}
The results of the preceding section may appear to be rather alarming: on the face of it, they seem to rule out a useful AdS/CFT description of thermal field theories [such as the one corresponding to the quark-gluon plasma] on spacetimes which are close to being flat [in the sense of being continuously deformable to a flat spacetime, unlike [say] the squashed Einstein static universe considered in Section 2 above]. For if we perform \emph{any} non-conformal deformation of the dual black hole with an exactly flat event horizon, then Seiberg-Witten pair-production will begin in the bulk, and this indicates a corresponding instability of field theories defined on spaces which are only slightly curved. However, there are two reasons not to jump to conclusions here.

First, one should bear in mind that an instability of such field theories is \emph{not} undesirable in all circumstances. In particular, one certainly does not want the holographic description of the quark-gluon plasma to work at all values of the temperature: for clearly such a plasma will not exist at arbitrarily low temperatures [see Figure 1]. Instead, the plasma always [that is, at all values of the chemical potential] passes through either a crossover or a phase transition as it is cooled. It is natural to interpret the phase change as the AdS/CFT dual of some instability developing in an AdS black hole spacetime as the electric charge is increased. In fact, it has been shown \cite{kn:AdSRN}\cite{kn:triple} that \emph{undistorted} black holes with flat event horizons do indeed become unstable to the Seiberg-Witten effect when they are sufficiently cold [at high values of the chemical potential\footnote{``High" here means ``well beyond the values associated with the quark matter critical point". Other effects, such as generalized Hawking-Page transitions, must dominate near the quark matter crossover.}], and it was proposed that this corresponds precisely to the fact that the quark-gluon plasma cannot exist at such values of the chemical potential and temperature. In short, the instability we have discussed here is not unwelcome at relatively low temperatures: on the contrary, the existence of some such effect is actually \emph{essential} if one is to arrive at a ``holographic" quark matter phase diagram. The fact that distortions of the underlying geometry render the system even more prone to instability is therefore not a cause for alarm, in this regime. The problem is to understand how instabilities can be avoided [only] at \emph{high} temperatures, where the quark-gluon plasma does indeed exist.

Secondly, it is not clear that this sort of instability will necessarily develop very quickly. If it develops unusually slowly for reasonable parameter choices, then the system might be ``effectively stable". Of course, in view of the comments above, one wants to ensure that this only happens at high temperatures.

There are in fact at least two effects that might slow Seiberg-Witten instability to the point where it becomes ineffective.

\bigskip

$\bullet$ Motivated by \cite{kn:DBI}, the authors of \cite{kn:albion} studied higher-dimensional versions of the BTZ black hole: that is, they performed a certain topological compactification of AdS. These holes have compact \emph{negatively} curved event horizons. These objects have negative scalar curvature at infinity, and so they are definitely unstable in the sense discussed here. However, a general argument is given in \cite{kn:albion} to the effect that the relevant eigenvalues become ``trapped" near zero, greatly slowing the evolution of the instability. More pertinently here, one can also consider massive black holes in AdS with compact negatively curved event horizons. These too are necessarily unstable; in fact, one can use the geometric techniques discussed in Section 3 to show that black holes with this topology always have negative scalar curvature at infinity, no matter how they are deformed. Again, however, it was shown in \cite{kn:barbon} that thermal effects \emph{at high temperatures} drastically slow the development of the instability. This is of course exactly what we need: the black hole is unstable, but this is physically relevant only at low temperatures, precisely where one expects the black hole description of the quark-gluon plasma to break down. It is very likely that the instability we have discussed here, which arises when a black hole with a flat event horizon is distorted, behaves in exactly the same way. That is, the field theory on a distorted flat spacetime \emph{is} actually unstable, but, at high temperatures, this effect is so slow as to be negligible.

\bigskip

$\bullet$ The brane action considered by Seiberg and Witten consists of two terms: a positive one contributed by the tension [and therefore related to the area of the brane] and a negative one contributed by the coupling to an antisymmetric tensor [which is proportional to the volume of the brane]. In the BPS case the leading order contributions of the two terms cancel, and the action depends very sensitively on the growth of the area and volume as a brane propagates towards infinity; this, as we discussed briefly in Section 2, is the reason for the ``fragility" of AdS black holes with spherical event horizons. In the case of a black hole with a \emph{flat} event horizon, the competition between the area and the volume is still more delicate, but, as we have seen, the action manages to remain positive if the event horizon is indeed perfectly flat. It is clear from Figure 2 that, whatever we do to such a black hole, the brane action remains positive near to the event horizon. However, as in the spherical case, any distortion of the black hole also distorts the bulk geometry around it; our results mean that, at a sufficient distance from the hole, this change allows the volume term to overcome the area term in the action. One might, however, find that the action will only become negative at some point \emph{which is very far from the black hole.} Causality then dictates that whatever happens in the region of negative action can only influence physics in the vicinity of the hole after the passage of a long period of time\footnote{``Time" here means ``time according to Killing observers", since this time is simply related to proper time at infinity. Of course, the time taken for any signal to \emph{reach} the event horizon is then infinite, which is why we speak of the ``vicinity" of the black hole. We circumvent this inessential complication by using the distance measured by these observers, this being finite in all cases.}. In the dual picture this would again correspond to an instability which develops extremely slowly. This effect is independent of the temperature, and therefore it is not really welcome: we \emph{want} Seiberg-Witten instability to set in rapidly at low temperatures, because such cold black holes correspond to an unphysical situation [a cold plasma] in the dual theory.

\bigskip

The first of these effects is discussed in detail in \cite{kn:barbon}, so we focus on the second: that is, let us try to assess whether bulk causality is likely to retard the development of Seiberg-Witten instability. The reader should bear in mind that it is our objective to show that this should \emph{not} happen.

If Q is the ADM charge, then the metric of a five-dimensional AdS Reissner-Nordstr$\ddot{\m{o}}$m black hole with a flat event horizon [generalizing the metric in equation (\ref{F})] is \cite{kn:lemmo}:
\begin{eqnarray}\label{K}
\m{g(\Lambda < 0;\;RN; \;T^3)} &=& \m{- \Bigg[{r^2\over L^2}\;-\;{2M\over
3\pi^2K^3r^2}\;+\;{Q^2\over 48\pi^5 K^6 r^4}\Bigg]dt^2} \\
&&\;+\;\m{{dr^2\over {r^2\over L^2}\;-\;{2M\over
3\pi^2K^3r^2}\;+\;{Q^2\over 48\pi^5 K^6 r^4}}}
\;+\; \m{r^2\Big[dx^2\;+\;dy^2\;+\;dz^2 \Big]}. \nonumber
\end{eqnarray}
This is the bulk geometry which is believed to be dual to a field theory which to some extent resembles the one describing a quark-gluon plasma \cite{kn:chamblin}.

When Q = 0, the BPS brane action in this geometry is never negative. If Q does not vanish, however, then, at a sufficient distance from the event horizon, it is possible for the brane action to become negative. In this sense, the AdS RN black hole resembles a distorted ``flat" black hole. There is a difference, however: this does not happen for all values of the charge. The action eventually becomes negative only if Q$^2$ exceeds a certain ``near extremal" value given [see \cite{kn:AdSRN}] by
\begin{equation}\label{L}
\m{{Q_{NE}^2\over L(KM)^{3/2}} \;=\;{16\pi^2 \over \sqrt{3}}}
\end{equation}
where L, as usual, is the AdS curvature scale; note that, in five dimensions, the expression on the left is indeed dimensionless. Note too that this number is smaller [by a factor of $\sqrt{27/32}$] than the number appearing in the definition of extremal charge in this case:
\begin{equation}\label{M}
\m{{Q_E^2\over L(KM)^{3/2}}\; = \; {64\sqrt{2}\pi^2 \over 9}.}
\end{equation}
Thus, there is a range of values of Q$^2$ between the near-extremal and the extremal values, such that the brane action eventually becomes negative. These are the black holes of interest to us here.

The action itself takes the form \cite{kn:AdSRN}
\begin{equation}\label{N}
\m{\$(\Lambda < 0;\;RN;\;T^3; M, L, Q, K, r) \;=\;16\pi^4\Theta P L^2 K^3\Bigg\{{{Q^2 \over 48\pi^5 K^6 r^2}\;-\;{2M \over 3\pi^2 K^3}
\over
1\;+\;\Bigg[\,1\;-\;{2 ML^2\over
3\pi^2 K^3r^4}\;+\;{Q^2L^2\over 48\pi^5 K^6 r^6}\,\Bigg]^{1/2}}\;+\;{r_{eh}^4\over L^2}\Bigg\}}.
\end{equation}
Here again the notation is as in equation (\ref{E}), apart from the presence of Q and of r$_{\m{eh}}$, the value of the radial coordinate at the event horizon. This is fixed by M, Q, K, and L as the largest real solution of the equation
\begin{equation}\label{O}
\m{{r_{eh}^6\over L^2}\;-\;{2Mr_{eh}^2\over
3\pi^2K^3}\;+\;{Q^2\over 48\pi^5 K^6} \;=\;0.}
\end{equation}

Our plan now is as follows. We shall consider black holes with perfectly flat event horizons but with a charge which slightly exceeds the ``near-extremal" value. The action of a BPS brane in this geometry is similar to that of such a brane in the spacetime around an initially flat but slightly distorted black hole: it is positive at first, but eventually becomes negative. In this geometry, it is straightforward to determine the distance from the event horizon to the point where the action becomes negative, and this could help us to judge whether causality is likely to have an appreciable effect on the rate at which Seiberg-Witten instability develops.

In detail, the procedure is as follows. We have to choose values of the parameters such that Q$^2$ slightly exceeds Q$_{\m{NE}}^2$, and then use equation (\ref{O}) to determine r$_{\m{eh}}$. This is then to be substituted into equation (\ref{N}); the smaller of the two roots of that expression will be r$_{\m{eh}}$, the larger, r = r$_{\$\,=\,0}$, will be the point beyond which the action becomes [and remains] negative. The distance between the two points, as seen by Killing observers at infinity, will be [with s = r/L]
\begin{equation}\label{P}
\m{d(r_{eh},\,r_{\$\,=\,0})\;=\;L\,\int^{s_{\$\,=\,0}}_{s_{eh}}{ds\over \Big[\,s^2\;-\;{2 M/L^2\over
3\pi^2 K^3 s^2}\;+\;{Q^2/L^4\over 48\pi^5 K^6 s^4}\,\Big]^{1/2}}}.
\end{equation}

The numerical results are as follows. First, we have verified that, for fixed values of all of the other parameters, $\m{d(r_{eh},\,r_{\$\,=\,0})}$ changes extraordinarily slowly as K and M are varied: see for example Figures 4 and 5, which portray the normalized action function $\$\$$(r) = $\m{\$(\Lambda < 0;\;RN;\;T^3; M, L, Q, K, r)/16\pi^4\Theta P L^2 K^3}$ for fixed values of all of the parameters except K, which changes from unity to 100.
\begin{figure}[!h] \centering
\includegraphics[width=0.9\textwidth]{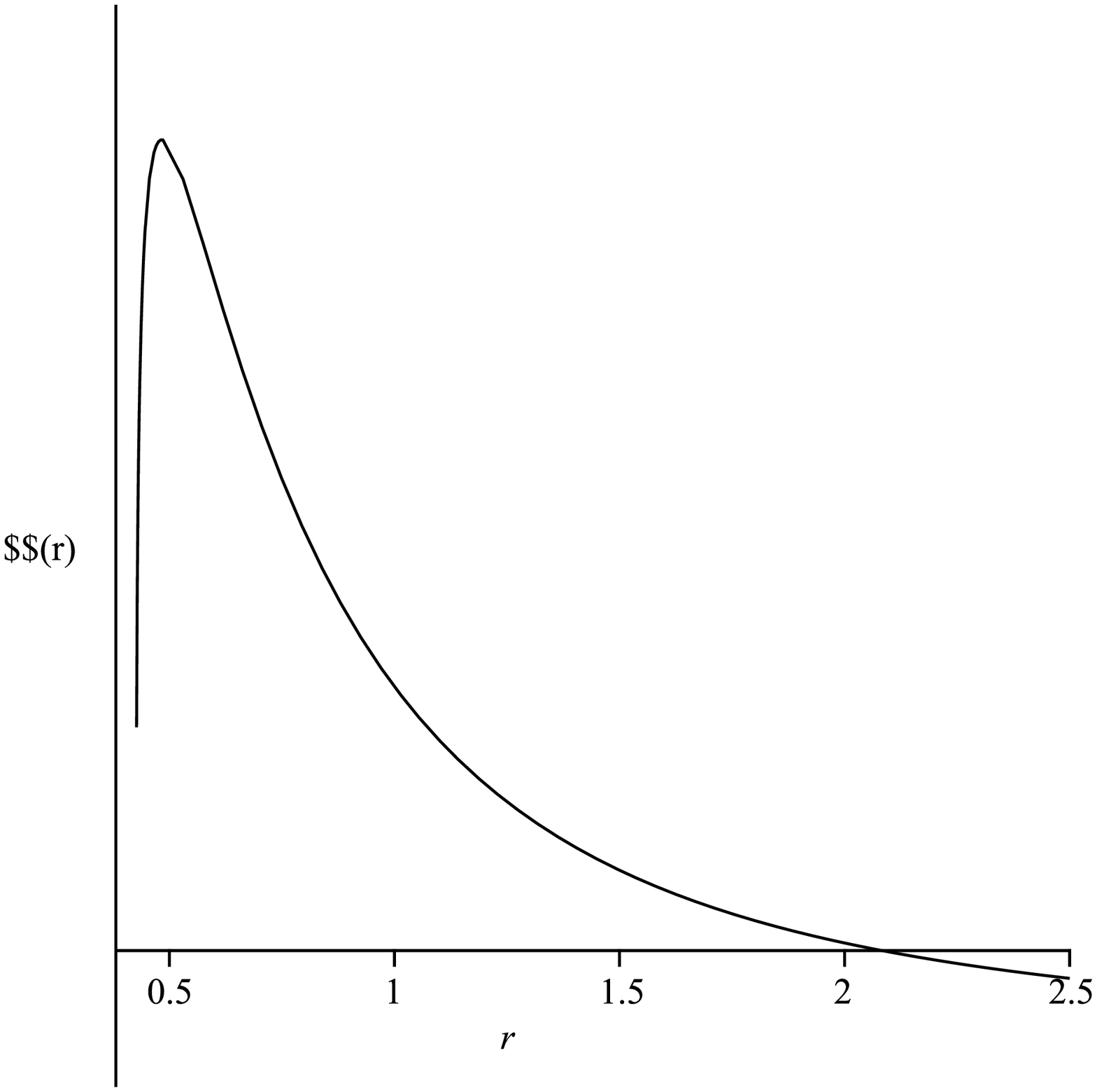}
\caption{\$\$(r) with $\m{Q^2/Q_{NE}^2 = 1.01, K=1, M/L^2=1, d(r_{eh},\,r_{\$\,=\,0})/L = 2.142689232}$.}
\end{figure}
\begin{figure}[!h] \centering
\includegraphics[width=0.9\textwidth]{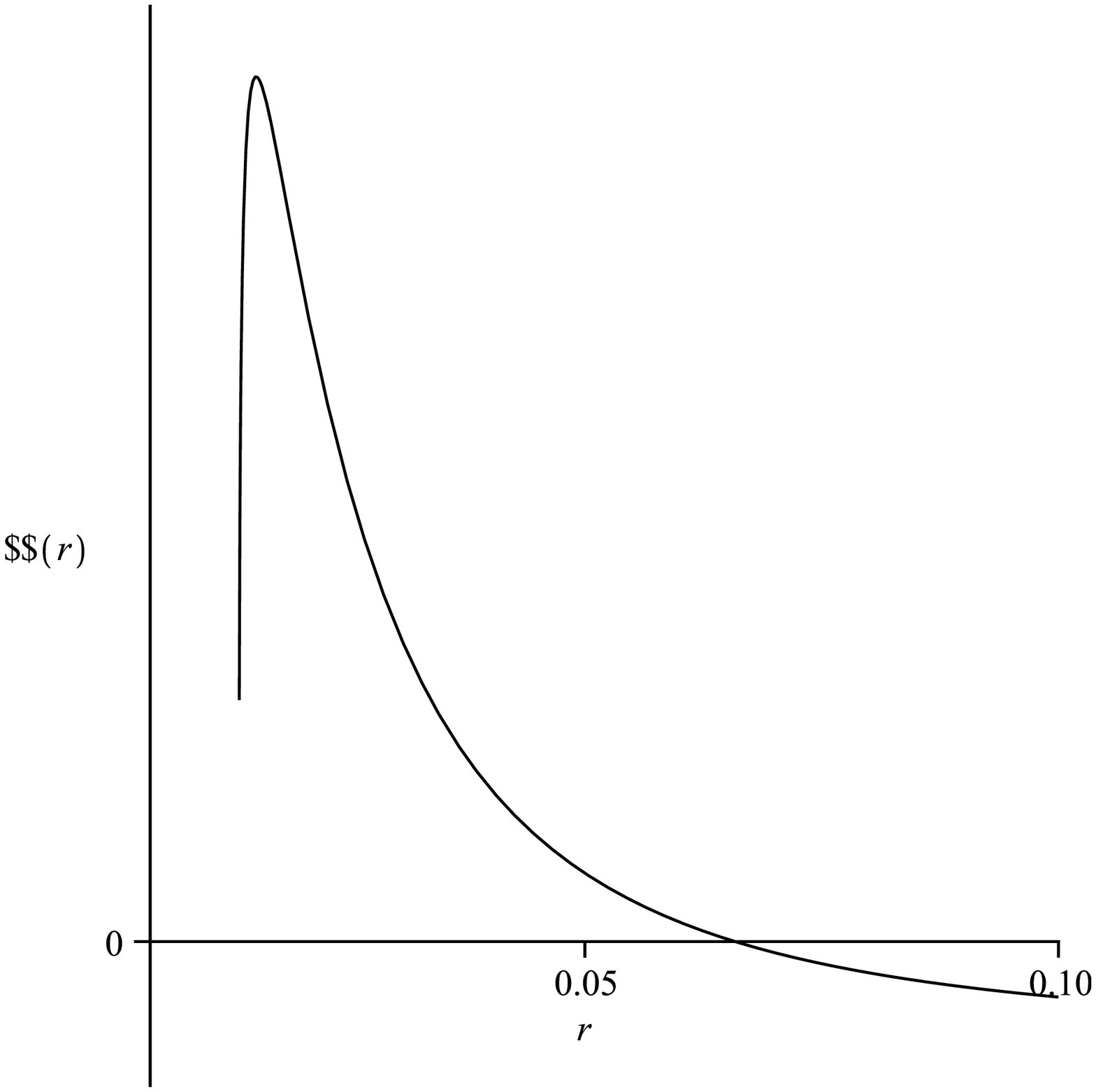}
\caption{\$\$(r) with $\m{Q^2/Q_{NE}^2 = 1.01, K=100, M/L^2=1, d(r_{eh},\,r_{\$\,=\,0})/L = 2.142689208}$.}
\end{figure}
We see that changing K does have a strong effect on r$_{\$\,=\,0}$; as K increases, it decreases [and so does r$_{\m{eh}}$]. However, this is of no significance in itself: r is an ``area coordinate", and the area of a section of the form r = constant at a fixed time is 8$\pi^3$K$^3$r$^3$. This quantity does increase with K: for large K, the event horizon has a large area, and the size of a brane at the point where the action begins to be negative is also large. But the distance from that brane to the horizon hardly changes. The situation as M changes is very similar. Therefore we fix K = M/L$^2$ = 1 henceforth.

The distance $\m{d(r_{eh},\,r_{\$\,=\,0})}$ does, as expected, depend on the quantity $\m{Q^2/Q_{NE}^2}$: as the charge approaches the ``near-extremal" value from above, r$_{\$\,=\,0}$ rapidly increases, so that the region of negative action recedes to greater distances from the black hole. However, as the table shows, this distance increases only extremely slowly.
\begin{center}
\begin{tabular}{|c|c|c|}
  \hline
$\m{Q^2/Q_{NE}^2}$ & $\m{r_{\$\,=\,0}/L}$  &  $\m{d(r_{eh},\,r_{\$\,=\,0})/L}$ \\
\hline
1.01 &    2.082941651      &  2.142689232   \\
1.001 &    6.759520013  &  3.304490376 \\
1.0001 &    21.42861654  &  4.456778510 \\
1.00001 &    67.78123140  &  5.608188336  \\
\hline
\end{tabular}
\end{center}
It is clear that bulk causality does not impose any significant constraints here. In other words, as the black hole is cooled, the presence of an infinite reservoir of negative free energy for brane pair-production will make itself felt almost immediately after the temperature drops below a certain critical value [which depends on the field-theoretic chemical potential]. We can conclude that cold [but still sub-extremal] AdS black holes with flat event horizons are genuinely unstable.

The black holes we have been considering in this section are all undistorted. As we have seen, the effect of distorting the event horizon is to induce Seiberg-Witten instability. \emph{Thus we see that the effect of distorting a cold AdS black hole with a flat event horizon is merely to exacerbate its pre-existing tendency to be unstable.} This means that the instability will actually occur earlier [that is, at higher temperatures] in the distorted case.

\addtocounter{section}{1}
\section* {\large{\textsf{5. Conclusion}}}
In summary, then, the combination of our results here with those of \cite{kn:AdSRN}\cite{kn:triple}\cite{kn:barbon} leads to the following picture. A thermal field theory on a spacetime which is not perfectly locally flat is unstable, since the dual system is a distorted black hole spacetime in which brane-antibrane pairs can nucleate in an uncontrolled way: it is a fragile black hole. This Seiberg-Witten instability is however a non-perturbative effect, since a barrier has to be surmounted. At high temperatures, the system is really metastable \cite{kn:barbon}, but this is not so at low temperatures, where in fact Seiberg-Witten instability is operative, and comes into effect quickly, even in the absence of spacetime curvature on the boundary. The presence of curvature only reinforces this process.
This is not unwelcome, because the dual to such a black hole is a plasma which indeed does not exist at low temperatures and high values of the chemical potential. In short, the fragility of these black holes helps us to construct a holographic account of the non-existence of cold quark plasmas, at all values of the chemical potential.

We conclude that black hole fragility is unlikely to be important in the cosmological application [to the QGP era of the early Universe, assuming [as is reasonable \cite{kn:lindetypical}] that the spatial sections are toral]. For while the spacetime in that case is far indeed from being flat, the temperatures involved are so extreme that the system is effectively stable.

In the region of Figure 1 just above the phase line BC, however, the situation is less clear-cut. In particular, near to the point B [the quark matter triple point], the temperature of the QGP is as low as possible; and such conditions could possibly arise in the region where a neutron star is being formed, where the spacetime curvature may be substantial. The role of fragility in this regime is to cause BC itself to be elevated higher in the diagram than one would otherwise expect: in other words, fragility means that spacetime curvature causes the QGP to be unstable even though the charge on the dual black hole is still below the ``near-extremal" value [equation (\ref{L})]. In short, \emph{the temperature of the triple point depends on the spacetime curvature}: all else being equal, it will be higher when the curvature is large.

Clearly much remains to be done to ensure that this picture is acceptable quantitatively. This will require a numerical investigation of black hole metrics of the form given in equation (\ref{H}) [or perhaps even of the form in equation (\ref{G})] which solve the field equations of some Gauss-Bonnet theory in the bulk, followed by an estimate [as in \cite{kn:triple}] of the temperature of the triple point. It would be particularly remarkable if the temperature around the triple point is strongly dependent on the spacetime curvature. At negligible curvatures, this region of the phase diagram may be probed experimentally in the foreseeable future: see for example \cite{kn:hohne}. On the other hand, the region around the triple point may ultimately be probed astrophysically, at high spacetime curvatures, as the process of formation of neutron stars comes to be better understood \cite{kn:alford2}. Both approaches might eventually lead to estimates of the temperature of the triple point; the present work suggests that the two results might not exactly coincide.

\addtocounter{section}{1}
\section*{\large{\textsf{Acknowledgement}}}
The author is grateful to Dr. Soon Wanmei for useful discussions.

\end{document}